\documentclass[aps,showpacs]{revtex4}
\usepackage{amsmath}
\usepackage{graphicx}

\begin{document}

\title{Emergent electrodynamics from the Nambu model for spontaneous Lorentz
symmetry breaking \\
}
\author{O. J. Franca$^{1}$, R. Montemayor$^{2}$ and L. F. Urrutia$^{1}$ }
\affiliation{$^{1}$ Instituto de Ciencias Nucleares, Universidad Nacional Aut{\'o}noma de
M{\'e}xico, A. Postal 70-543, 04510 M{\'e}xico D.F., M{\'e}xico }
\affiliation{$^{2}$ Instituto Balseiro and CAB, Universidad Nacional de Cuyo and CNEA,
8400 Bariloche, Argentina}

\begin{abstract}
After imposing the Gauss law constraint as an initial condition upon the
Hilbert space of the Nambu model, in all its generic realizations, we
recover QED in the corresponding non-linear gauge $A_{\mu }A^{\mu
}=n^{2}M^{2}$. Our result is non-perturbative in the parameter $M$ for $%
n^{2}\neq 0$ and can be extended to the $n^{2}=0$ case.\textbf{\ }This shows
that in the Nambu model, spontaneous Lorentz symmetry breaking dynamically
generates gauge invariance, provided the Gauss law is imposed as an
initial condition. In this way electrodynamics is recovered, with the photon
being realized as the Nambu-Goldstone modes of the spontaneously broken
symmetry, which finally turns out to be non-observable
\end{abstract}

\maketitle

\section{Introduction}

The possible violation of Lorentz invariance has recently received a lot of
attention, both from the experimental and the theoretical perspective. In
this latter case, the interest lies mainly in connection with possible
effects arising from the drastic modifications of space-time at distances of
the order of Planck length, suggested by most of current quantum gravity
approaches. Experiments and astrophysical observations put severe bounds
upon the parameters describing such violations, which are still being
improved.

In this work we revisit a different and interesting idea related to
spontaneous Lorentz symmetry breaking, which is that photons (and gravitons)
could emerge as the corresponding Nambu-Goldstone bosons from that process.
We focus on the case of electrodynamics. Normally the null mass of both
particles is explained by invoking some type of gauge invariance. Since this
almost sacred principle has undoubtedly been fundamental in the development
of physics, it is very interesting to explore the possibility that it could
have a dynamical origin\cite{PRLNielsen}. This idea goes back to the works
of Nambu \cite{Nambu} and Bjorken \cite{Bjorken} together with many other
contributions \cite{others}. Recently it has been revived in references \cite%
{kostelecky1,chkareulli, azatov,mohapatra,chkareulli1,BLUHM}.

One of the most explored approaches starts form a theory with a vector field
$B^{\mu }$ endowed with the standard electrodynamics kinetic term plus a
potential designed to break the Lorentz symmetry via a non-zero vacuum
expectation value $<B^{\mu }>$, which defines a preferred direction $n_{\mu
} $ in space-time. This potential also breaks gauge invariance. The original
model was proposed by Nambu \cite{Nambu} and recently has been generalized
to incorporate the so called bumblebee models \cite{kostelecky2}. In
general, the subsequent symmetry breaking obtained from the non-zero minimum
of the potential, splits the original four degrees of freedom $B^{\mu }$
into three vectorial Nambu-Goldstone bosons $A^{\mu }$, to be identified
with the photon satisfying the constraint $A_{\mu }A^{\mu }= n^{2}M^{2}$,
plus a massive scalar field $\sigma $, which is assumed to be excited at
very high energies.

It is noteworthy to recall here that the constraint $A_{\mu }A^{\mu }=\pm
M^{2}$ was originally proposed by Dirac as a way to derive the
electromagnetic current from the additional excitations of the photon field,
which now has ceased to be gauge degrees of freedom, avoiding in this way
the problems arising from considering point-like charges \cite{Diracfull}.
As reported in \cite{Dirac1}, this theory requires the existence of an ether
emerging from quantum fluctuations, but does not necessarily imply the
violation of Lorentz invariance, due to an averaging process of the quantum
states producing such fluctuations.

In the framework of the Nambu model, calculations at the tree level \cite%
{Nambu} and to the one loop level \cite{azatov} have been carried out,
showing that the possible Lorentz violating effects are not present in the
physical observables and that the results are completely consistent with the
standard gauge invariant electrodynamics. This is certainly a surprising
result, which can be truly appreciated from the complexity of the
calculation in Ref.\cite{azatov} and that certainly deserves additional
understanding. The above mentioned calculations have been performed using an
expansion of the non-linear interaction up to order $1/M$ and $1/M^{2}$,
respectively. The basic question that remains is to which extent the
non-perturbative Nambu model describes the standard effects of full QED,
which have been experimentally measured with astonishing precision.

In this work we recover and extend to the space-like and\ light-like\textbf{%
\ }cases the results regarding the equivalence of the time-like version ($%
n_{\mu }n^{\mu }>0$) of the Nambu model and QED, given in Ref. \cite{BLUHM}.
Our proposal is to start with the full analysis of these theories at the
classical level and then move on to the quantum regime. Even at the
classical level the problem is a difficult one because of the presence of
the non-linear constraint $A_{\mu }A^{\mu }=n^{2}M^{2}$, although non-linear
gauges have been successfully used in the quantization of Yang-Mills
theories \cite{joglekar}. Despite this, and selecting an approach which does
not require to completely fix the gauge, we show that the Nambu model, with
the Gauss law as an additional constraint, imposed as an initial condition,
has exactly the same canonical structure of the electromagnetic field. This
point becomes evident when both canonical structures are expressed in terms
of the electric $\mathbf{E}$ and potential $\mathbf{A}$ fields. Since the
quantum dynamics is determined by the canonical structure, this result
warrants that both theories, in their non-perturbative form, are equivalent
at the quantum level, thus yielding the same physical results.

The paper is organized as follows. In Section II we review the Nambu model,
showing that its space-like version arises from the spontaneous Lorentz
symmetry breaking (SLSB) of a bumblebee model. The remaining sectors of the
Nambu model are considered as a natural generalization of the space-like
case, without specifying the mechanism leading to the SLSB\textbf{.} In
Section III we show that taking the phase space of the Nambu model and
imposing the Gauss law constraint on it as an initial condition, the
resulting canonical structure coincides with that of the electromagnetic
field in the nonlinear gauge given by the particular SLSB considered. In
Section IV we show that the separate procedures for dealing with the Nambu
model, presented in Section III, can be given a unified description using a
Lagrange multiplier to impose the resolution of the non-linear constraint in
each case. Finally, Section V contains the conclusions and final comments.
Appendix A summarizes the canonical version of QED, which we use as a
benchmark to state the equivalence with the different realizations of the
Nambu model plus the Gauss law. Appendix B includes a specific example of
the construction of the potentials in the non-relativistic version of the
non-linear gauge, for the case of a constant magnetic field. Finally, in
Appendix C we discuss the dynamics of the full SLSB bumblebee model in its
space-like version and show how electrodynamics is recovered in this case%
\textbf{.}

\section{SLSB and the Nambu model}

As we will show in the following, the Nambu model can be motivated as the
low energy limit of \textbf{\ }the bumblebee model (the metric is $\eta
_{\mu \nu }=diag(1,-1,-1,-1)$)%
\begin{equation}
\mathcal{L}(B_{\mu })=-\frac{1}{4}B_{\mu \nu }B^{\mu \nu }-\frac{m^{2}}{2}%
\left( B_{\mu }B^{\mu }\right) -\frac{\lambda }{4}\left( B_{\mu }B^{\mu
}\right) ^{2}-J_{\mu }B^{\mu },\;\;\;\partial _{\mu }J^{\mu }=0,
\label{LAGNAMBU0}
\end{equation}%
with
\begin{equation}
B_{\mu \nu }=\partial _{\mu }B_{\nu }-\partial _{\nu }B_{\mu },
\end{equation}%
after the $SO(1,3)$ symmetry is broken. Following the standard steps we
obtain the broken symmetry Lagrangian%
\begin{eqnarray}
\mathcal{L} &=&-\frac{1}{4}F_{\mu \nu }F^{\mu \nu }\left( 1+\frac{\omega }{%
\sigma _{0}}\right) ^{2}+\frac{1}{\sigma _{0}}\left( 1+\frac{\omega }{\sigma
_{0}}\right) F_{\mu \nu }\left( A^{\mu }\partial ^{\nu }\omega \right)
\notag \\
&&-\frac{1}{2}n^{2}\left( \partial _{\mu }\omega \right) \left( \partial
^{\mu }\omega \right) +\frac{1}{2\sigma _{0}^{2}}A^{\nu }A^{\mu }\left(
\partial _{\mu }\omega \right) \left( \partial _{\nu }\omega \right)   \notag
\\
&&-\frac{1}{2}\left( -2m^{2}n^{2}\right) \omega ^{2}-\lambda \sigma
_{0}\omega ^{3}-\frac{\lambda }{4}\omega ^{4}-J_{\mu }A^{\mu }\left( 1+\frac{%
\omega }{\sigma _{0}}\right) ,  \label{LAG4}
\end{eqnarray}%
where we have successively set $B^{\mu }=\Lambda _{\;\;\nu }^{\mu }(x)n^{\nu
}\sigma (x)\equiv \frac{1}{\sigma _{0}}A^{\mu }(x)\sigma (x)$,$\;\sigma
(x)=\sigma _{0}+\omega (x)$, and $F_{\mu \nu }=\partial _{\mu }A_{\nu
}-\partial _{\nu }A_{\mu }$. This parameterization \ requires
\begin{equation}
B_{\mu }B^{\mu }=\sigma ^{2}n^{2},\;\;A_{\mu }A^{\mu }=\sigma _{0}^{2}n^{2}.
\label{CONSPAR}
\end{equation}%
Here $n^{\nu }\;$is a constant vector characterizing the symmetry breaking
vacuum according to
\begin{eqnarray}
n^{\alpha }\;\mathrm{time-like} &:&SO(1,3)\rightarrow SO(3),  \label{TEMP} \\
n^{\alpha }\;\mathrm{space-like} &:&SO(1,3)\rightarrow SO(1,2).  \label{ESP}
\end{eqnarray}%
Let us emphasize that the light-like case ($n^{2}=0$) is forbidden because
then $B_{\mu }B^{\mu }=0$, so that there is no spontaneous symmetry breaking
since the potential turns out to be zero. The Goldstone modes are contained
in the field $A^{\mu }$,\ which is restricted by the condition (\ref{CONSPAR}%
), so that it contains only three degrees of freedom. The field $\omega (x)\;
$describes the massive fluctuations of $\sigma (x)$ around the minimum $%
\sigma _{0}$ of the potential, where%
\begin{equation}
M_{\omega }=\sqrt{2m^{2}}.
\end{equation}%
In this way the parameter $m^{2}\;$in (\ref{LAGNAMBU0}) must be positive. We
also require $\lambda >0$, in such a way that the potential is bounded at
infinity. The minimum of the potential has to be real and satisfies
\begin{equation}
\sigma _{0}=\sqrt{-\frac{m^{2}n^{2}}{\lambda }}\;>0,
\end{equation}%
which imposes $n^{2}<0.\;$Let us further consider the case where the
effective theory described by (\ref{LAG4}) is in the energy range $%
(E<<M_{\omega })$ where the field $\omega (x)$ is not excited. In this
situation we have%
\begin{equation}
\mathcal{L}=-\frac{1}{4}F_{\mu \nu }F^{\mu \nu }-J_{\mu }A^{\mu
},\;\;\;\;\;\;\;A_{\mu }A^{\mu }=-\frac{m^{2}}{\lambda }<0.  \label{EMERGQED}
\end{equation}%
It is interesting to remark that the choice $n^{2}<0\;$in (\ref{LAG4})\ also
produces the right sign normalization for the kinetic term of the field $%
\omega .$

In this way we have shown that the spontaneous Lorentz symmetry breaking of
the bumblebee model (\ref{LAGNAMBU0}) leads to the Nambu model \cite{Nambu}
only in its space-like version. We consider the remaining sectors of the\
Nambu model as an extension of this result, without deriving them from the
spontaneous symmetry breaking \ of a specific model. Thus we replace the
second equation in (\ref{EMERGQED}) by the requirement\textbf{\ }%
\begin{equation}
A_{\mu }A^{\mu }=M^{2}n^{2},\;\;\;M^{2}>0,  \label{EXTENSION}
\end{equation}%
where $n^{2}$ can now be positive, negative or zero.\ In each case we will
show that $n^{\mu }\;$indeed describes the vacuum around which the symmetry
has been broken by the Nambu-Goldstone modes\ $A_{\mu }$.

The standard analysis of the model is made by explicitly solving the
constraint $A_{\mu }A^{\mu }=M^{2}n^{2}$ and subsequently substituting the
result into the Lagrangian (\ref{EMERGQED}), thus providing a theory
analogous to the standard non-linear sigma model\cite{NLSIGMAM}. As we will
show in Section IV, an alternative compact way of dealing with the model is
to introduce the constraint via a Lagrange multiplier.

In the following we examine the conditions under which this theory is
equivalent to standard electrodynamics, in such a way that $A_{\mu }$ can be
interpreted as the photon field in the non-linear gauge $A_{\mu }A^{\mu
}=M^{2}n^{2}$. The choices of $n^{2}$ provide alternative forms in which the
constraint $A_{\mu }A^{\mu }=M^{2}n^{2}\;$is to be solved, in such a way
that the remaining symmetry is manifest. For the time-like case we have
\begin{equation}
A_{0}=\sqrt{M^{2}+\mathbf{A}^{2}},  \label{CTL}
\end{equation}%
with $\mathbf{A}=(A^{1},A^{2},A^{3})\;$being three independent degrees of
freedom, and where the variable $A_{0}$\ is explicitly invariant under
rotations.

The situation in the space-like case is
\begin{equation}
A_{3}=\sqrt{M^{2}+\left( A^{0}\right) ^{2}-\left( A^{1}\right) ^{2}-\left(
A^{2}\right) ^{2}},  \label{CSL}
\end{equation}%
where the independent variables are $(A^{0},A^{1},A^{2})\;$and the variable $%
A_{3}$ is explicitly invariant under the corresponding $SO(1,2)\;$subgroup.

In the case $n^{2}=0$ we introduce the following parameterization\ \cite%
{Nambu}
\begin{equation}
A_{0}=c\left( 1+\frac{B}{c}+\frac{A_{1}^{2}+A_{2}^{2}}{4c^{2}(1+\frac{B}{c})}%
\right) ,\quad A_{3}=c\left( 1+\frac{B}{c}-\frac{A_{1}^{2}+A_{2}^{2}}{%
4c^{2}(1+\frac{B}{c})}\right) ,  \label{NAMBUPARAM}
\end{equation}%
which provides the expansion parameter $c$\ for\textbf{\ }the chosen
dependent fields\textbf{, }which is analogous to $M$\ in the remaining
cases. Here the independent variables are $B(x),\;A_{1}(x)\;$and $A_{2}(x)$.

We identify the vacuum $\left( A_{\mu }\right) _{0}\;$of each sector of the
theory by setting equal to zero all space-time dependent fields in $A_{\mu }$%
. In this way we obtain%
\begin{eqnarray}
n^{2} &>&0,\;\;\left( A_{\mu }\right) _{0}=(M,0,0,0),  \notag \\
n^{2} &<&0,\;\;\left( A_{\mu }\right) _{0}=(0,0,0,M),  \notag \\
n^{2} &=&0,\;\;\left( A_{\mu }\right) _{0}=(c,0,0,c).  \label{VACNAMBU}
\end{eqnarray}

\section{Electrodynamics as a constrained Nambu model}

The first hint upon how the Nambu model is related to electrodynamics arises
from the Lagrangian equations of motion. The variation of (\ref{EMERGQED})\
produces%
\begin{equation}
\delta \mathcal{L}=\left( \partial _{\mu }F^{\mu \nu }-J^{\nu }\right)
\delta A_{\nu }.  \label{LAGNAMBU}
\end{equation}%
and the equations of motion for each case are obtained looking at the
variation of the independent variables
\begin{eqnarray}
\mathrm{time-like\ case} &:&\delta A_{0}=\frac{A_{i}}{A_{0}}\delta A_{i},
\label{TL1} \\
\mathrm{space-like\ case} &:&\delta A_{3}=\frac{A_{0}}{A_{3}}\delta A_{0}-%
\frac{A_{a}}{A_{3}}\delta A_{a},  \label{SL1}
\end{eqnarray}%
which produces
\begin{equation}
\left( \partial _{k}F^{k0}-J^{0}\right) \frac{A_{i}}{A_{0}}+\left( \partial
_{\mu }F^{\mu i}-J^{i}\right) =0,\;\;i=1,2,3,  \label{TL2}
\end{equation}%
or%
\begin{eqnarray}
\left( \partial _{k}F^{k0}-J^{0}\right) +\left( \partial _{\mu }F^{\mu
3}-J^{3}\right) \frac{A_{0}}{A_{3}} &=&0,  \notag \\
\left( \partial _{\mu }F^{\mu a}-J^{a}\right) -\left( \partial _{\mu }F^{\mu
3}-J^{3}\right) \frac{A_{a}}{A_{3}} &=&0,\;\;a=1,2.  \label{SL22}
\end{eqnarray}%
respectively. Thus we see that in order to recover electrodynamics from the
Nambu model we must further impose the Gauss law $\partial
_{i}F^{i0}-J^{0}=0\;$as an additional constraint. This cannot be done at the
Lagrangian level because such a constraint involves the velocity $\dot{A}%
_{i}.$

Recalling that the quantum dynamics is determined by the canonical
structure, our general strategy to prove the equivalence between the
quantized Nambu model plus the Gauss law and QED will be to perform the
Hamiltonian analysis of the former and to compare with the corresponding
formulation of electrodynamics, summarized in Appendix A.

\subsection{The time-like case}

We start from the Lagragian density (\ref{EMERGQED}), where we have
explicitly solved and substituted the non-linear constraint in the form $%
A_{0}=\sqrt{\mathbf{A}^{2}+M^{2}}$,%
\begin{equation}
\mathcal{L}_{N}(A_{i})=\frac{1}{2}\left( \left( \mathbf{\partial }_{t}A_{i}%
\mathbf{-}\partial _{i}\left( \sqrt{\mathbf{A}^{2}+M^{2}}\right) \right)
^{2}-\left( \mathbf{\nabla \times A}\right) ^{2}\right) -J_{0}\sqrt{\mathbf{A%
}^{2}+M^{2}}-J_{i}A^{i}.  \label{LAGtime-like}
\end{equation}%
The canonically conjugated momenta $\Pi _{i}$ are%
\begin{equation}
\Pi _{i}=\frac{\partial L}{\partial \dot{A}_{i}}=\partial _{0}A_{i}-\partial
_{i}\sqrt{M^{2}+\mathbf{A}^{2}}=E_{i},
\end{equation}%
leading to the canonical Hamiltonian density $\mathcal{H}=E_{i}\dot{A}_{i}-%
\mathcal{L}$%
\begin{equation}
\mathcal{H}=\frac{1}{2}\mathbf{E}^{2}+\frac{1}{2}\mathbf{B}%
^{2}+E_{i}\partial _{i}\sqrt{M^{2}+\mathbf{A}^{2}}+J_{0}\sqrt{\mathbf{A}%
^{2}+M^{2}}+J_{i}A^{i},  \label{NAMBUHAM}
\end{equation}%
with the basic non-zero Poisson brackets%
\begin{equation}
\left\{ A_{i}(\mathbf{x},t),\;E_{j}(\mathbf{y},t)\right\} =\delta
_{ij}\delta ^{3}(\mathbf{x}-\mathbf{y}).  \label{ORIGINALB}
\end{equation}%
The Hamiltonian equations of motion\ turn out to be%
\begin{equation}
-\dot{E}_{i}+\left( \partial _{k}E_{k}-J_{0}\right) \frac{A_{i}}{\sqrt{M^{2}+%
\mathbf{A}^{2}}}+\left[ \nabla \times B\right] _{i}-J^{i}=0,  \label{EQH1}
\end{equation}%
where%
\begin{equation}
E_{i}=\dot{A}_{i}-\partial _{i}\sqrt{M^{2}+\mathbf{A}^{2}}%
,\;\;B_{i}=\epsilon _{ijk}\partial _{j}A^{k}.
\end{equation}%
In other words, we have recovered the Lagrangian equations of motion (\ref%
{TL2})\ for the Nambu model in this representation.

The next step is to implement the Gauss law constraint via an arbitrary
function\ $\chi $, so that the modified Hamiltonian density is%
\begin{equation}
\mathcal{\tilde{H}}=\frac{1}{2}\mathbf{E}^{2}+\frac{1}{2}\mathbf{B}%
^{2}-\left( \partial _{i}E_{i}-J_{0}\right) \left( \sqrt{M^{2}+\mathbf{A}^{2}%
}+\mathbf{\chi }\right) +J_{i}A^{i}.  \label{MODHAMDEN}
\end{equation}%
Redefining the arbitrary function%
\begin{equation}
\Theta =\mathbf{\chi }+\sqrt{M^{2}+\mathbf{A}^{2}},
\end{equation}%
we finally obtain%
\begin{equation}
\mathcal{\tilde{H}}=\frac{1}{2}\left( \mathbf{E}^{2}+\mathbf{B}^{2}\right)
-\Theta \left( \partial _{i}E_{i}-J_{0}\right) +J_{i}A^{i},
\end{equation}%
with the corresponding first order action%
\begin{equation}
S=\int d^{3}y\left( E_{i}\dot{A}_{i}-\mathcal{\tilde{H}}\right) .
\end{equation}%
In this way, the Hamiltonian density adopts the form (\ref{HAMEDF1}) of the
Appendix A. The Poisson bracket relations also coincide with those of (\ref%
{DBED}). We have thus shown that the imposition of the Gauss law into the
Nambu Hamiltonian density (\ref{NAMBUHAM}), which is performed in (\ref%
{MODHAMDEN}), does in fact lead to classical electrodynamics. Let us remark
that, as will be shown in Section IV, the dynamics of the Nambu model, for
any choice of $n^{2}$, preserves the Gauss law in time, so that it is
sufficient to impose it as an initial condition.

\subsection{The space-like case}

Here we start from \ (\ref{EMERGQED}), but making the substitution $A_{3}=%
\sqrt{M^{2}+A_{0}^{2}-A_{a}A_{a}}$%
\begin{equation}
\mathcal{L}_{N}(A^{0},A^{a})=\frac{1}{2}\left( \mathbf{E}_{T}^{2}+E_{3}^{2}-%
\mathbf{B}^{2}\right) -J_{0}A^{0}-J^{a}A_{a}-J^{3}\sqrt{%
M^{2}+A_{0}^{2}-A_{a}A_{a}},\quad a=1,2,
\end{equation}%
where the independent degrees of freedom are now $A_{0},A_{a}$. Since $%
A_{3}\;$is just a shorthand for$\;\sqrt{M^{2}+A_{0}^{2}-A_{T}^{2}}$, we have
\begin{equation}
\dot{A}_{3}=\frac{A_{0}}{A_{3}}\dot{A}_{0}-\frac{A_{a}}{A_{3}}\dot{A}%
_{a},\;\;\;a=1,2,
\end{equation}%
which leads to%
\begin{eqnarray}
E_{3} &=&\frac{A_{0}}{A_{3}}\dot{A}_{0}-\frac{A_{a}}{A_{3}}\dot{A}%
_{a}-\partial _{3}A_{0},  \label{E3} \\
E_{a} &=&\dot{A}_{a}-\partial _{a}A_{0}.  \label{EA}
\end{eqnarray}%
The canonically conjugated momenta are%
\begin{eqnarray}
\Pi _{0} &=&\frac{\partial L_{N}}{\partial \dot{A}_{0}}=E_{3}\frac{A_{0}}{%
A_{3}}=\left( \frac{A_{0}}{A_{3}}\dot{A}_{0}-\frac{A_{a}}{A_{3}}\dot{A}%
_{a}-\partial _{3}A_{0}\right) \frac{A_{0}}{A_{3}},  \label{PI0} \\
\Pi _{a} &=&\frac{\partial L_{N}}{\partial \dot{A}_{a}}=E_{a}-E_{3}\frac{%
A_{a}}{A_{3}}=\dot{A}_{a}-\partial _{a}A_{0}-\left( \frac{A_{0}}{A_{3}}\dot{A%
}_{0}-\frac{A_{b}}{A_{3}}\dot{A}_{b}-\partial _{3}A_{0}\right) \frac{A_{a}}{%
A_{3}},  \label{PIA}
\end{eqnarray}%
with the non-zero Poisson brackets now being%
\begin{equation}
\left\{ A_{0}(\mathbf{x},t),\Pi _{0}(\mathbf{y},t)\right\} =\delta ^{3}(%
\mathbf{x}-\mathbf{y}),\;\;\;\left\{ A_{a}(\mathbf{x},t),\;\Pi _{b}(\mathbf{y%
},t)\right\} =\delta _{ab}\delta ^{3}(\mathbf{x}-\mathbf{y}).  \label{PBSL}
\end{equation}%
Solving for the velocities in (\ref{PI0}) and (\ref{PIA}) yields
\begin{equation}
\dot{A}_{0}=\left( \frac{\mathbf{A}_{T}^{2}}{A_{0}^{2}}\Pi _{0}+\frac{1}{%
A_{0}}\left( A_{a}\Pi _{a}+A_{a}\partial _{a}A_{0}+A_{3}\partial
_{3}A_{0}\right) \right) ,  \label{VEL0}
\end{equation}%
\begin{equation}
\dot{A}_{a}=\Pi _{a}+\partial _{a}A_{0}+\frac{A_{a}}{A_{0}^{2}}A_{0}\Pi _{0}.
\label{VELa}
\end{equation}%
We can now write the electric field in terms of the momenta as follows%
\begin{equation}
E_{a}=\Pi _{a}+A_{a}\frac{\Pi _{0}}{A_{0}},\;\;\;E_{3}=A_{3}\frac{\Pi _{0}}{%
A_{0}}.  \label{EFIELD}
\end{equation}%
The Hamiltonian density%
\begin{equation}
\mathcal{H}=\Pi _{0}\dot{A}_{0}+\Pi _{a}\dot{A}_{a}-\frac{1}{2}\left(
\mathbf{E}_{T}^{2}+E_{3}^{2}-\mathbf{B}^{2}\right)
+J^{0}A_{0}+J^{a}A_{a}+J^{3}\sqrt{M^{2}+A_{0}^{2}-A_{a}A_{a}}
\end{equation}%
can be written as%
\begin{eqnarray}
\mathcal{H} &=&\frac{1}{2}\frac{\Pi _{0}^{2}}{A_{0}^{2}}\left(
A_{a}A_{a}+A_{3}^{2}\right) +\frac{1}{2}\Pi _{a}\Pi _{a}+\frac{\Pi _{0}}{%
A_{0}}A_{a}\Pi _{a}  \notag \\
&&-A_{0}\left( \partial _{a}\left( \frac{\Pi _{0}}{A_{0}}A_{a}\right)
+\partial _{3}\left( \frac{\Pi _{0}}{A_{0}}A_{3}\right) +\partial _{a}\Pi
_{a}\right)   \notag \\
&&+\frac{1}{2}\mathbf{B}^{2}(A_{0},A_{a})+J^{0}A_{0}+J^{a}A_{a}+J^{3}\sqrt{%
M^{2}+A_{0}^{2}-A_{a}A_{a}}.  \label{HAMSL1}
\end{eqnarray}%
To make contact with the standard form of the Hamiltonian in QED we rewrite%
\begin{equation}
\frac{1}{2}\Pi _{a}\Pi _{a}+\frac{\Pi _{0}}{A_{0}}A_{a}\Pi _{a}=\frac{1}{2}%
E_{a}^{2}-\frac{1}{2}\left( \frac{\Pi _{0}}{A_{0}}\right) ^{2}A_{a}A_{a},
\label{REL1}
\end{equation}%
using (\ref{EFIELD}), and realize that the Gauss law adopts the following
form in terms of the independent variables
\begin{equation}
\partial _{i}E_{i}=\partial _{a}E_{a}+\partial _{3}E_{3}=\partial _{a}\Pi
_{a}+\partial _{a}\left( A_{a}\frac{\Pi _{0}}{A_{0}}\right) +\partial
_{3}\left( A_{3}\frac{\Pi _{0}}{A_{0}}\right) .  \label{REL2}
\end{equation}%
Substituting (\ref{REL1}) and (\ref{REL2}) in (\ref{HAMSL1}) we obtain
\begin{equation}
\mathcal{H}=\frac{1}{2}E_{3}^{2}+\frac{1}{2}E_{a}^{2}+\frac{1}{2}\mathbf{B}%
^{2}(A_{0},A_{a})-A_{0}\left( \partial _{i}E_{i}-J_{0}\right)
+J^{a}A_{a}+J^{3}A_{3}.  \label{HAMNAMBU}
\end{equation}%
The above Hamiltonian density corresponds to the Nambu model and the
resulting equations of motion coincide with those obtained in (\ref{SL22}).
Eq. (\ref{HAMNAMBU}) has just the form of the standard Hamiltonian density
for electrodynamics, except that $A_{3},E_{3}$ are notations for the
corresponding functions in terms of the dynamical coordinates and momenta.
Notice that in this case $A_{0}\;$is not a Lagrange multiplier, so that we
are still missing the Gauss law to recover the electrodynamics.

Next, we follow the established procedure by imposing the Gauss law to the
Nambu Hamiltonian (\ref{HAMNAMBU}) via the arbitrary function$\;\chi $. The
result is%
\begin{eqnarray}
\mathcal{\tilde{H}} &=&\frac{1}{2}E_{3}^{2}+\frac{1}{2}E_{a}^{2}+\frac{1}{2}%
\mathbf{B}^{2}(A_{0},A_{a})  \notag \\
&&-\left( A_{0}+\chi \right) \left( \partial _{i}E_{i}-J_{0}\right)
+J^{a}A_{a}+J^{3}A_{3},
\end{eqnarray}%
The redefinition $\Theta =A_{0}+\chi \;$leaves%
\begin{equation}
\mathcal{\tilde{H}}=\frac{1}{2}\left( \mathbf{E}^{2}+\mathbf{B}^{2}\right)
-\Theta \left( \partial _{i}E_{i}-J_{0}\right) +J^{a}A_{a}+J^{3}A_{3},
\label{HAMSL:}
\end{equation}%
which is precisely of the form (\ref{HAMEDF1}). As emphasized at the end of
Subsection A, it is enough to impose the Gauss law as an initial condition.
Now we use the transformations from the original space-like variables $%
A_{0},A_{a},\;\Pi _{0},\Pi _{a}$ to the set $A_{i},E_{j}\;(i,j=1,2,3).$%
\begin{eqnarray}
A_{a} &=&A_{a},\;\;\;\;\;\;\;\;\;\;\;\;\;\;\;A_{3}=\sqrt{%
M^{2}+A_{0}^{2}-A_{a}A_{a}},  \label{DEF1} \\
E_{a} &=&\Pi _{a}+A_{a}\frac{\Pi _{0}}{A_{0}},\;\;\;\;E_{3}=\frac{\Pi _{0}}{%
A_{0}}\sqrt{M^{2}+A_{0}^{2}-A_{a}A_{a}},  \label{DEF2}
\end{eqnarray}%
together with the brackets (\ref{PBSL}) for $A_{0},A_{a},\Pi _{0},\Pi _{a}$,
to verify that we recover the brackets among $A_{i},E_{j}\;$given in (\ref%
{DBED}), which define classical electrodynamics. In other words, we have
shown that the change of variables (\ref{DEF1}) , (\ref{DEF2})\ is a
canonical transformation. We also have%
\begin{equation}
\Pi _{0}\dot{A}_{0}+\Pi _{a}\dot{A}_{a}=E_{i}\dot{A}_{i},
\end{equation}%
as expected.

As a final comment in this section we observe that the possible
singularities envisaged in Section IV of Ref. \cite{BLUHM}, arising in the
space-like case due to the possibility of $A_{0}=0$\ and which are not
present in the time-like case, do in fact appear in our construction via the
factor $1/A_{0}\;$which occurs in many of the intermediate steps in this
Section, and also in the corresponding equations in Section IV.
Nevertheless, such singularity is absent both in the final Hamiltonian and
in the final Dirac brackets.\textbf{\ }

\subsection{The light-like case}

In this case the degrees of freedom are the fields $B$ and $A^{a}$, $a=1,2$,
according to (\ref{NAMBUPARAM}). To simplify the expressions we redefine $%
c+B\rightarrow B$, and thus we get%
\begin{align}
A_{0}& =\frac{1}{4B}\left( 4B^{2}+A_{a}A_{a}\right) ,  \label{a0ll} \\
A_{3}& =\frac{1}{4B}\left( 4B^{2}-A_{a}A_{a}\right) .  \label{a3ll}
\end{align}%
From here on  $A_{0}$ and $A_{3}$ are just notation for the functions given by the above  equations,
and their time derivatives are%
\begin{align}
\dot{A}_{0}& =\frac{1}{B}\left( A_{3}\dot{B}+\frac{1}{2}A_{a}\dot{A}%
_{a}\right) , \\
\dot{A}_{3}& =\frac{1}{B}\left( A_{0}\dot{B}-\frac{1}{2}A_{a}\dot{A}%
_{a}\right) .
\end{align}%
In terms of these degrees of freedom, the components of the electric field
are
\begin{align}
E_{3}& =\dot{A}_{3}-\partial _{3}A_{0}=\frac{1}{B}\left( A_{0}\dot{B}-\frac{1%
}{2}A_{a}\dot{A}_{a}-B\partial _{3}A_{0}\right) , \label{E3LL}\\
E_{a}& =\dot{A}_{a}-\partial _{a}A_{0}.\label{EaLL}
\end{align}%
As in the preceding cases, the starting point is the Lagrangian density (\ref%
{EMERGQED}), which takes the form%
\begin{align}
\mathcal{L}_{N}(A^{0},A^{a})& =\frac{1}{2}\left( \mathbf{E}%
_{a}^{2}+E_{3}^{2}-\mathbf{B}^{2}\right)   \notag \\
& -J_{0}\left( B+\frac{A_{a}^{2}}{4B}\right) -J^{3}\left( B-\frac{A_{a}^{2}}{%
4B}\right) -J^{a}A_{a}.
\end{align}

The canonical momenta of $A_{a}$ and $B$ can be written
\begin{equation}
\Pi _{a}=\frac{\partial \mathcal{L}}{\partial \dot{A}_{a}}=E_{a}-\frac{A_{a}%
}{2B}E_{3},\ \ \ \ \ \ \ \Pi _{B}=\frac{\partial \mathcal{L}}{\partial \dot{B%
}}=\frac{A_{0}}{B}E_{3},  \label{2}
\end{equation}%
and satisfy the non-zero fundamental Poisson brackets%
\begin{equation}
\left\{ B\left( \mathbf{x},t\right) ,\Pi _{B}\left( \mathbf{y},t\right)
\right\} =\delta ^{3}\left( \mathbf{x}-\mathbf{y}\right) ,\ \ \ \ \ \ \ \ \
\ \left\{ A_{a}\left( \mathbf{x},t\right) ,\Pi _{b}\left( \mathbf{y}%
,t\right) \right\} =\delta _{ab}\delta ^{3}\left( \mathbf{x}-\mathbf{y}%
\right) .  \label{LLPB}
\end{equation}%
From (\ref{2}), the electric field in terms of the momenta is%
\begin{equation}
E_{a}=\Pi _{a}+\frac{A_{a}}{2A_{0}}\Pi _{B},\ \ \ \ \ \ \ E_{3}=\frac{B}{%
A_{0}}\Pi _{B}.  \label{3}
\end{equation}%
and the velocities can be expressed%
\begin{align}
\dot{A}_{a}& =\Pi _{a}+\partial _{a}A_{0}+\frac{A_{a}}{2A_{0}}\Pi _{B}, \\
\dot{B}& =\frac{1}{A_{0}}\left( B\Pi _{B}+B\partial _{3}A_{0}+\frac{1}{2}%
A_{a}\left( \Pi _{a}+\partial _{a}A_{0}\right) \right) .
\end{align}%
The Hamiltonian density
\begin{equation}
\mathcal{H}=\Pi _{a}\dot{A}_{a}+\Pi _{B}\dot{B}-\frac{1}{2}\left( \mathbf{E}%
_{a}^{2}+E_{3}^{2}-\mathbf{B}^{2}\right) +J_{0}\left( B+\frac{A_{a}^{2}}{4B}%
\right) +J^{3}\left( B-\frac{A_{a}^{2}}{4B}\right) +J^{a}A_{a}
\end{equation}%
takes now the form%
\begin{align}
\mathcal{H}& =\frac{1}{2}\Pi _{a}^{2}+\frac{B}{2A_{0}}\Pi _{B}^{2}+\frac{%
A_{a}}{2A_{0}}\Pi _{B}\Pi _{a}+\frac{1}{2}\mathbf{B}^{2}  \notag \\
& +\left[ \partial _{a}\Pi _{a}+\partial _{3}\left( \frac{\Pi _{B}B}{A_{0}}%
\right) +\partial _{a}\left( \frac{A_{a}}{2A_{0}}\Pi _{B}\right) +J^{0}%
\right] A_{0}+J^{3}A_{3}+J^{a}A_{a}.  \label{HAMLL}
\end{align}%
Introducing the electric and magnetic field according to (\ref{3}), we have%
\begin{equation}
\mathbf{E}^{2}=\frac{B}{A_{0}}\left( \Pi _{B}\right) ^{2}+\left( \Pi
_{a}\right) ^{2}+\frac{A_{a}}{A_{0}}\Pi _{B}\Pi _{A_{a}},
\end{equation}%
and the Gauss law becomes%
\begin{align}
\partial _{i}F^{i0}-J^{0}& =\partial _{3}E^{3}+\partial _{a}E^{a}-J^{0}
\notag \\
& =-\left[ \partial _{3}\left( \frac{B}{A_{0}}\Pi _{B}\right) +\partial
_{a}\left( \Pi _{A_{a}}+\frac{A_{a}}{2A_{0}}\Pi _{B}\right) +J^{0}\right] .
\end{align}%
These relations imply that the Hamiltonian can be expressed, using the
electric and magnetic fields and the Gauss law,  in the compact form%
\begin{equation}
\mathcal{H}=\frac{1}{2}\left( \mathbf{E}^{2}+\mathbf{B}^{2}\right) -\left(
\partial _{i}F^{i0}-J^{0}\right) A^{0}+J_{a}A^{a}+J_{3}A^{3},
\end{equation}%
which corresponds again to the Nambu model. As in the previous cases,
imposing the Gauss law as an initial constraint, we obtain the Hamiltonian
of the electromagnetic field in the nonlinear gauge $A_{\mu }A^{\mu }=0$%
\begin{equation}
\mathcal{H}=\frac{1}{2}\left( \mathbf{E}^{2}+\mathbf{B}^{2}\right) -\left(
\partial _{i}F^{i0}-J^{0}\right) \Theta +J_{i}A^{i}
\end{equation}%
where $\Theta $ is an arbitrary function, corresponding to the primary
constraint $\left( \partial _{i}F^{i0}-J^{0}\right) $. Similarly to the
space-like case, in the intermediate steps of this construction it is
possible for singularities to appear because of the factor $1/B$. However,
as in the space-like case, such singularities are not present in the final
Hamiltonian density and in the Dirac brackets. Once more, starting from the
definitions (\ref{a0ll}-\ref{a3ll}) together with (\ref{E3LL}-\ref{EaLL}) in
terms of the canonical variables, and using the brackets (\ref{LLPB}) it is
a direct matter to recover the Poisson algebra of electrodynamics (\ref{DBED}%
) in terms of the fields $A_{i}$ and $E_{i}$.

\section{Unified discussion of the Nambu model}

The alternative possibilities to realize the spontaneous symmetry breaking
can be jointly discussed by incorporating the non-linear constraint via a
Lagrange multiplier $\lambda $%
\begin{equation}
\mathcal{L}=-\frac{1}{4}F_{\mu \nu }F^{\mu \nu }-J_{\mu }A^{\mu }+\lambda
(A_{\mu }A^{\mu }-M^{2}n^{2}),  \label{a0}
\end{equation}%
which yields the equations of motion%
\begin{align}
\partial _{\mu }F^{\mu \nu }-J^{\nu }+2\lambda A^{\nu }& =0,  \label{a00} \\
A_{\mu }A^{\mu }-M^{2}n^{2}& =0.  \label{a01}
\end{align}%
From Eq. (\ref{a0}) we obtain the relation%
\begin{equation}
\partial _{0}\left( \partial _{i}F^{i0}-J^{0}\right) =-\partial _{k}\left(
\frac{A^{k}}{A^{0}}\left( \partial _{i}F^{i0}-J^{0}\right) \right) ,
\label{GLCONS}
\end{equation}%
which implies that the imposition of the Gauss law as an initial condition
guarantees that $\partial _{i}F^{i0}-J^{0}=0$ remains valid for arbitrary
times. It is also important to emphasize that the choice $\lambda (x,t=0)=0$%
\ as an initial condition leads, according to Eq. (\ref{a0}), to the
requirement that all four Maxwell equations must be satisfied at $t=0$, in
particular the\ Gauss law.

There are several possibilities to work out $\lambda (x,t)$. If we take a
particular value for the free index in Eq.(\ref{a00}), $\nu =\sigma $, to
solve the Lagrange multiplier, it yields
\begin{equation}
\lambda =-\frac{1}{2A^{\sigma }}\left( \partial _{\mu }F^{\mu \sigma
}-J^{\sigma }\right) ,
\end{equation}%
(no sum over $\sigma $) and thus we obtain%
\begin{align}
A^{\sigma }\left( \partial _{\mu }F^{\mu \nu }-J^{\nu }\right) -\left(
\partial _{i}F^{i\sigma }-J^{\sigma }\right) A^{\nu }& =0,  \label{b1} \\
A_{\mu }A^{\mu }-M^{2}n^{2}& =0.  \label{b2}
\end{align}%
For $\sigma =0$ these equations correspond to the time-like case (\ref{TL2}%
), and for $\sigma =3$ they produce the space-like equations of motion (\ref%
{SL22}). This shows that the Lagrange multiplier formulation of the Nambu
model is equivalent to the direct-substitution approach developed in Section
III. If we introduce the definitions
\begin{equation}
E_{i}=\dot{A}_{i}-\partial _{i}A_{0},\;\;\;\;B_{i}=\epsilon _{ijk}\partial
_{j}A^{k},
\end{equation}%
and impose an additional constraint, the Gauss law $\mathbf{\nabla }\cdot
\mathbf{E}-J^{0}=0$, the above equations become Maxwell equations in the
nonlinear gauge (\ref{b2}). Let us emphasize that (\ref{b1}) contains the
two cases corresponding to (\ref{CTL}) and (\ref{CSL}),\emph{\ }together
with the light-like choice $n^{2}=0$.

The quantum dynamics of this model is determined by the canonical formalism.
For the Nambu model such a formulation can be obtained in a very compact
way, directly from the Lagrangian density (\ref{a0}), using the Dirac
approach for singular Lagrangians. The coordinate fields are $A^{\mu }$ and $%
\lambda $, and the definitions of their canonical momenta leads to%
\begin{equation}
\pi _{i}=\frac{\partial \mathcal{L}}{\partial \dot{A}^{i}}=F^{0i},\qquad \pi
_{0}=\frac{\partial \mathcal{L}}{\partial \dot{A}^{0}}=0,\qquad \pi
_{\lambda }=\frac{\partial \mathcal{L}}{\partial \dot{\lambda}}=0.
\label{a3}
\end{equation}%
From here we can make explicit the time derivative of the spatial components
of the vector field%
\begin{equation}
\dot{A}^{i}=-\pi ^{i}+\partial ^{i}A^{0},
\end{equation}%
and two primary constraints emerge%
\begin{equation}
\phi _{1}=\pi _{0},\qquad \phi _{2}=\pi _{\lambda }.  \label{a6}
\end{equation}%
Thus, the extended Hamiltonian density results%
\begin{equation}
\mathcal{H}=-\frac{1}{2}\pi _{i}\pi ^{i}+\frac{1}{4}F_{ij}F^{ij}+J_{i}A^{i}+%
\left( J_{0}-\partial _{i}\pi ^{i}\right) A^{0}-\lambda \left( A_{\mu
}A^{\mu }-M^{2}n^{2}\right) +u\pi _{0}+v\pi _{\lambda }.
\end{equation}%
The consistency conditions of the primary constraints lead to the secondary
ones%
\begin{align}
\phi _{3}& =\left\{ \pi _{\lambda },H\right\} =A_{\mu }A^{\mu }-M^{2}n^{2}=0,
\\
\phi _{4}& =\left\{ \pi _{0},H\right\} =2\lambda A_{0}+(\partial _{i}\pi
^{i}-J_{0})=0.
\end{align}%
In turn, the equations of motion of these last constraints fix the arbitrary
functions $u$ and $v$. Thus we have the set of second class constraints $%
\phi _{1}$, $\phi _{2}$, $\phi _{3}$, and $\phi _{4}$, which determine a
subspace where the Hamiltonian density becomes%
\begin{equation}
H=-\frac{1}{2}\pi _{i}\pi ^{i}+\frac{1}{4}F_{ij}F^{ij}+\left( J_{0}-\partial
_{i}\pi ^{i}\right) A^{0}+j_{i}A^{i}.  \label{a17}
\end{equation}%
To make explicit the dynamics it is also necessary to construct the Dirac
brackets that define the symplectic structure of the model. The constraint $%
\phi _{3}$ admits several realizations. One of these expresses the time
component of $A^{\mu }$ in terms of the spatial ones (the time-like case)
\begin{equation}
A_{0}=\sqrt{M^{2}-A^{i}A_{i}}.  \label{a20}
\end{equation}%
Here the spatial components of $A^{\mu }$ are the degrees of freedom, and
the resulting Dirac brackets are%
\begin{equation}
\left\{ A^{j}\left( \mathbf{x},t\right) ,\pi _{i}\left( \mathbf{y},t\right)
\right\} _{D}=\eta _{i}^{j}\delta ^{3}\left( \mathbf{x-y}\right) .
\label{a21}
\end{equation}%
Then, the canonical momenta conjugated to the $A^{j}$ fields are directly
the $\pi _{j}$. Another realization is obtained when we take a spatial
component as the dependent field (the space-like case), such as%
\begin{equation}
A_{3}=\left( M^{2}+A^{0}A_{0}+A^{a}A_{a}\right) ^{1/2},  \label{a23}
\end{equation}%
where $a=1,2$. Now the Dirac brackets among the degrees of freedom $\left(
A^{0},A^{1},A^{2}\right) $ and the momenta $\pi _{j}$ become%
\begin{align}
\left\{ A_{0}\left( \mathbf{x},t\right) ,\pi _{3}\left( \mathbf{y},t\right)
\right\} _{D}& =\frac{1}{A_{0}\left( \mathbf{x},t\right) }\left(
M^{2}+A_{0}^{2}\left( \mathbf{x},t\right) +A^{a}\left( \mathbf{x},t\right)
A_{a}\left( \mathbf{x},t\right) \right) ^{1/2}\,\delta ^{3}\left( \mathbf{x-y%
}\right) ,  \label{a24} \\
\left\{ A_{0}\left( \mathbf{x},t\right) ,\pi _{a}\left( \mathbf{y},t\right)
\right\} _{D}& =-\frac{A_{a}\left( \mathbf{x},t\right) }{A_{0}\left( \mathbf{%
x},t\right) }\delta ^{3}\left( \mathbf{x-y}\right) ,  \label{a25} \\
\left\{ A^{a}\left( \mathbf{x},t\right) ,\pi _{3}\left( \mathbf{y},t\right)
\right\} _{D}& =0,  \label{a26} \\
\left\{ A^{a}\left( \mathbf{x},t\right) ,\pi _{b}\left( \mathbf{y},t\right)
\right\} _{D}& =\eta _{b}^{a}\delta ^{3}\left( \mathbf{x-y}\right) .
\label{a27}
\end{align}%
The momenta $\pi _{j}$ are not the canonical conjugate momenta of the
degrees of freedom $\left( A^{0},A^{1},A^{2}\right) $, but from the above
Dirac brackets we obtain%
\begin{equation}
\left\{ A^{3}\left( \mathbf{x},t\right) ,\pi _{3}\left( \mathbf{y},t\right)
\right\} _{D}=\left\{ \left( M^{2}+A^{0}\left( \mathbf{x},t\right)
A_{0}\left( \mathbf{x},t\right) +A^{a}\left( \mathbf{x},t\right) A_{a}\left(
\mathbf{x},t\right) \right) ^{1/2},\pi _{3}\left( \mathbf{y},t\right)
\right\} _{D}=\delta ^{3}\left( \mathbf{x-y}\right) ,
\end{equation}%
and thus the $A^{i}$ and $\pi _{j}$ fields satisfy the same Dirac brackets
as in the time-like case%
\begin{equation}
\left\{ A^{i}\left( \mathbf{x},t\right) ,\pi _{j}\left( \mathbf{y},t\right)
\right\} _{D}=\eta _{j}^{i}\delta ^{3}\left( \mathbf{x-y}\right) .
\end{equation}%
From the first equation in (\ref{a3}) we have $\pi _{i}=-E_{i}$ in both
cases. Imposing the Gauss law
\begin{equation}
\left( \partial _{i}E_{i}-J^{0}\right) =0,  \label{a18}
\end{equation}%
as an initial constraint on the phase space of the Nambu model, we obtain
the extended Hamiltonian density for the electromagnetic field%
\begin{equation}
H=\frac{1}{2}\left( \mathbf{E}^{2}+\mathbf{B}^{2}\right) -\mathbf{J}\cdot
\mathbf{A}-\Theta \left( \mathbf{\nabla }\cdot \mathbf{E}-J^{0}\right) ,
\end{equation}%
where $\Theta $ is an arbitrary function, with the usual brackets between
the potential and the electric field%
\begin{equation}
\left\{ A_{i}\left( \mathbf{x},t\right) ,E_{j}\left( \mathbf{y},t\right)
\right\} =\delta _{ij}\delta ^{3}\left( \mathbf{x-y}\right) .
\end{equation}

The constrained phase space of the Hamiltonian for the free Nambu model
includes subspaces where (\ref{a17}) becomes negative, because of the term $%
A^{0}\partial _{i}\pi ^{i}$. In general the existence of these subspaces
leads to instabilities and spoils unitarity. When the Gauss law is imposed
as an additional constraint to obtain electrodynamics in the non linear
gauge, this term disappears from the constrained Hamiltonian, which becomes
positive definite and yields a unitary theory.

\section{FINAL COMMENTS}

The Nambu model can be motivated by the spontaneous Lorentz symmetry
breaking in the space-like sector of the bumblebee model given by Eq. (\ref%
{LAGNAMBU0}). It is defined by the Lagrangian density of electrodynamics
plus the non-linear constraint $A_{\mu }A^{\mu }=n^{2}M^{2}$ and, in this
sense, it is similar to the non-linear sigma model. The standard way of
dealing with the Nambu model is by solving the constraints in their
different choices according to the character of $n_{\mu }$, which now is
arbitrary\textbf{, }and then substituting the solution into the Lagrangian.
In this way, it is clear that the Nambu model contains three massless
degrees of freedom, corresponding to the Nambu-Goldstone modes of the SLSB,
and manifestly breaks gauge invariance. As shown by Nambu, at the classical
and tree level \cite{Nambu}, this model plus the Gauss law as an additional
constraint, is equivalent to standard electrodynamics in the non-linear
gauge $A_{\mu }A^{\mu }=n^{2}M^{2}$. Further calculations including one loop
effects \cite{azatov} corroborate this conjecture. The above mentioned
calculations are carried only up to order $1/M$ and $1/M^{2}$, respectively.
On the other hand, Ref. \cite{BLUHM} contains a discussion of the
non-perturbative equivalence in the time-like case.

Comparing with quantum electrodynamics a la Dirac, as described in Appendix
A, where the Gauss law is imposed as a first class constraint upon the
Hilbert space of the problem, we are able to show that the quantized Nambu
model in the time-like and space-like cases, plus the quantized Gauss law
imposed on the corresponding Hilbert space does indeed reproduce QED in a
non-perturbative way with respect to the parameter $M$. This result is also
extended to the light-like case ($n^{2}=0$). It is important to emphasize
that the dynamics of the Nambu model guarantees the conservation of the
Gauss law, in such a way that it is sufficient to impose it as an initial
condition. This is the main result of this work.

The proof goes as follows. For each case (time-like, space-like or
light-like) we start from the corresponding canonical degrees of freedom,
arising after solving the constraint $A_{\mu }A^{\mu }=n^{2}M^{2}$, and
obtain their Dirac brackets. After rewriting the standard electrodynamic
variables $A_{i},E_{i}$ in terms of the canonical degrees of freedom and
with the use of the related Dirac brackets, we prove that they satisfy the
symplectic algebra of electrodynamics, given by the brackets (\ref{DBED})%
\textbf{.} Moreover, each of the original Hamiltonian densities (\ref%
{NAMBUHAM}), (\ref{HAMSL1}) and (\ref{HAMLL}) adopt the form (\ref%
{HAMEDF1}) of electrodynamics, after the Gauss law is imposed via an
arbitrary function upon them. Since the quantum dynamics is
determined by the canonical structure, we can now construct the
corresponding quantum formalism. The fields $A_{i},E_{i}$ will satisfy the
commutation relations of QED, arising from the brackets (\ref{DBED}), and
the respective Hamiltonian densities will become the QED Hamiltonian after
the Gauss law is imposed. Since this constraint is first class, we adopt
here Dirac's point of view that the quantum theory is defined in the Hilbert
subspace of physical states that are annihilated by the constraint. The fact
that QED is recovered in the gauge corresponding to each solution of the
non-linear constraint can be traced back to the particular initial form of
the Nambu model used in each case.

It is appropriate to remark that the imposition of the Gauss law in the
otherwise non-gauge invariant Nambu model, is equivalent to restoring such
invariance upon the full system, because this first class constraint is the
generator of gauge transformations. For this reason a more accurate
statement is that gauge invariance is indeed dynamically recovered from the
Nambu model in order to reproduce QED, provided it is imposed as an initial
condition. This statement coincides with the results in Refs. \cite%
{Nambu,BLUHM}. The results of Ref. \cite{azatov} also support this point of
view because the perturbative calculation performed there is carried on a
Hilbert space where the photons are transverse to begin with, and therefore
the Gauss law is imposed from the outset.

As shown in Section IV, the separate procedures described in Section III can
be given a unified description using a Lagrange multiplier to impose the
resolution of the non-linear constraint in each case. In this way it is
straightforward to show that for every realization of a spontaneous Lorentz
symmetry breaking consistent with $A_{\mu }A^{\mu }=n^{2}M^{2}$, and after
imposing the Gauss law, we obtain electrodynamics in the corresponding
non-linear gauge. Furthermore, using the Dirac method, we explicitly show
that in the canonical formalism for the time-like and space-like cases, the
canonical brackets among the electric and the potential field are the
same as in electrodynamics.

Nambu has also proposed that the solution of the associated Hamilton-Jacobi
equation for a particle in a given electromagnetic field should provide, via
the principal Hamilton Jacobi function, the function that allows to make the
change of gauge from an initial one (here taken as the Coulomb gauge) to the
non-linear gauge $A_{\mu }A^{\mu }=n^2 M^{2}$ \cite{Nambu}. In Appendix B we
have explicitly constructed an example of this procedure for the case of a
constant magnetic field. We have considered the $1/M$ (with $M\rightarrow
\infty $) limit of the non-linear constraint in the time-like case, which is
consistent with the non-relativistic limit of the Hamilton-Jacobi equation.

Finally we discuss the full SLSB model, corresponding to the space-like
case, arising from the bumblebee model (\ref{LAGNAMBU0}), where the massive
field becomes dynamical instead of being automatically frozen as is done in
the Nambu model. The equations of motion again preserve the time evolution
of the Gauss law. Its imposition as an initial condition produces a
consistent solution for the massive field which leads to electrodynamics in
the related space-like non-linear gauge.

\appendix

\section{Electrodynamics}

Let us briefly review the standard Hamiltonian formulation of
electrodynamics. We start from the second order action%
\begin{equation}
S_{2}(A_{i},A_{0})=\int d^{3}y\;\mathcal{L}_{2}=\int d^{3}y\left[ \frac{1}{2}%
\left( \left( \dot{A}_{i}\mathbf{-}\partial _{i}A_{0}\right) ^{2}-\left(
\mathbf{\nabla \times A}\right) ^{2}\right) -J^{0}A_{0}-J^{i}A_{i}\right] .
\end{equation}%
The canonical momenta are
\begin{equation}
\Pi _{i}=\frac{\partial \mathcal{L}}{\partial \dot{A}_{i}}=\dot{A}_{i}%
\mathbf{-}\partial _{i}A_{0}=E_{i},\;\ \Pi _{0}=0,
\end{equation}%
which produce the Hamiltonian density%
\begin{equation}
\mathcal{H=}\frac{1}{2}\mathbf{\Pi }^{2}+\frac{1}{2}\left( \mathbf{\nabla
\times A}\right) ^{2}\mathcal{-}\left( \partial _{i}E_{i}-J^{0}\right)
A_{0}+J^{i}A_{i}.  \label{HAMED}
\end{equation}%
We also have the PB%
\begin{eqnarray}
\left\{ A_{i}(\mathbf{x},t),A_{j}(\mathbf{y},t)\right\} &=&0,\;\ \ \ \ \ \ \
\ \ \ \ \ \ \ \ \ \ \ \ \ \left\{ E_{i}(\mathbf{x},t),E_{i}(\mathbf{y}%
,t)\right\} =0, \\
\left\{ A_{i}(\mathbf{x},t),\Pi _{j}(\mathbf{y},t)\right\} &=&\delta
_{ij}\delta ^{3}(\mathbf{x}-\mathbf{y}),\;\;\left\{ A_{0}(\mathbf{x},t),\Pi
_{0}(\mathbf{y},t)\right\} =\delta ^{3}(\mathbf{x}-\mathbf{y}).  \label{PBED}
\end{eqnarray}%
The canonical theory is constructed via Dirac's method, due to the fact that
the constraint\ $\ \Sigma =\Pi _{0}\simeq 0$ is present.

The extended Hamiltonian density is
\begin{equation}
\mathcal{H}_{E}=\frac{1}{2}\mathbf{E}^{2}+\frac{1}{2}\left( \mathbf{\nabla
\times A}\right) ^{2}\mathcal{-}A_{0}\left( \partial _{i}E_{i}-J^{0}\right)
+J^{i}A_{i}+u\Pi _{0}.
\end{equation}%
The condition%
\begin{equation}
0=\dot{\Sigma}(x)=\left\{ \Sigma (x),\int d^{3}y\;\mathcal{H}_{E}(y)\right\}
,
\end{equation}%
leads to the Gauss law constraint
\begin{equation}
\Omega =\left( \partial _{i}E_{i}-J^{0}\right) .
\end{equation}%
Finally we obtain$\;\dot{\Omega}=0$, in virtue of current conservation. In
this way, the final first order action for electrodynamics is
\begin{equation}
S_{1}(A_{i},E_{i},A_{0})=\int d^{3}y\;\mathcal{L}_{1}=\int d^{3}y\;\left(
E_{i}\dot{A}_{i}-\mathcal{H}\right) ,  \label{ACTIONED}
\end{equation}%
where%
\begin{equation}
\mathcal{H=}\frac{1}{2}(\mathbf{E}^{2}+\mathbf{B}^{2})+J^{i}A_{i}-A_{0}%
\left( \partial _{i}E_{i}-J^{0}\right) +u\Pi _{0},  \label{HAMEDF}
\end{equation}%
with the two first class constraints%
\begin{equation}
\Pi _{0}\simeq 0,\;\partial _{i}E_{i}-J^{0}\simeq 0.  \label{EDCONSTRAINTS}
\end{equation}%
Normally one fixes
\begin{equation}
\Pi _{0}\simeq 0,\;\;\;A_{0}\simeq \Theta
\end{equation}%
with $\Theta \;$an arbitrary function to be consistently determined after
the remaining first class constraint is fixed, which yields
\begin{equation}
\mathcal{H=}\frac{1}{2}(\mathbf{E}^{2}+\mathbf{B}^{2})+J^{i}A_{i}-\Theta
\left( \partial _{i}E_{i}-J^{0}\right) .  \label{HAMEDF1}
\end{equation}%
The remaining PB are%
\begin{equation}
\left\{ A_{i}(\mathbf{x},t),A_{j}(\mathbf{y},t)\right\} =0,\;\left\{ E_{i}(%
\mathbf{x},t),E_{i}(\mathbf{y},t)\right\} =0,\;\left\{ A_{i}(\mathbf{x}%
,t),E_{j}(\mathbf{y},t)\right\} =\delta _{ij}\delta ^{3}(\mathbf{x}-\mathbf{y%
}).  \label{DBED}
\end{equation}%
The equations of motion are%
\begin{equation}
\dot{A}_{i}=\left\{ A_{i},H\right\} ,\;\rightarrow \mathbf{E}=-\mathbf{\dot{A%
}}-\mathbf{\nabla }\Theta .
\end{equation}%
Taking $\mathbf{\nabla \times \;}$from the above equation we recover
Faraday's law%
\begin{equation}
\mathbf{\nabla \times E}=-\mathbf{\nabla \times \dot{A}}=-\mathbf{\dot{B}.}
\end{equation}%
The remaining equation arises from%
\begin{equation}
\dot{E}_{i}=\left\{ E_{i},H\right\} ,\;\rightarrow \;\mathbf{\dot{E}-\nabla
\times B=J}.
\end{equation}

\section{Explicit gauge transformation in the non-relativistic case}

In general, it proves to be very difficult to go from a known gauge to the
non-linear gauge $A_{\mu }A^{\mu }=M^{2}$. Let us assume that we start from
the Coulomb gauge with potentials $\tilde{A}_{\mu }$, then the required
potentials in the non linear gauge are obtained by the transformation%
\begin{equation}
A_{\mu }=\tilde{A}_{\mu }+\partial _{\mu }\Lambda ,
\end{equation}%
together with the equation%
\begin{equation}
\left( \tilde{A}_{\mu }+\partial _{\mu }\Lambda \right) ^{2}=M^{2},
\end{equation}%
which determines the required gauge function $\Lambda .$

As suggested by Nambu, $\Lambda $ can be identified with the principal
Hamilton-Jacobi function $S$ corresponding to a particle with charge $q$ and
mass $m$ moving in the corresponding electromagnetic fields. In fact, the
related Hamilton-Jacobi equation is%
\begin{equation}
\left( \partial _{\mu }S-\frac{q}{c}A_{\mu }\right) ^{2}=m^{2}c^{2},
\end{equation}%
in such a way that the relation between the two problems is%
\begin{equation}
\Lambda =-\frac{c}{q}S,\;\;M=\frac{c^{2}m}{q}.
\end{equation}%
Here we give an example of such a construction for the time-like case, in
the non-relativistic approximation. The non-relativistic Hamilton-Jacobi
equation is%
\begin{equation}
\left( \partial _{t}\bar{S}-q\Phi \right) =-\frac{1}{2m}\left( \partial _{i}%
\bar{S}+\frac{q}{c}A^{i}\right) ^{2},  \label{NONRELIMHJ}
\end{equation}%
where%
\begin{equation}
\bar{S}=S\mp mc^{2}t.
\end{equation}%
The minus sign choice in the square root arises from the corresponding form
of the Hamilton-Jacobi equation for the free particle%
\begin{equation}
\partial _{t}\bar{S}+\frac{1}{2m}\left( \partial _{i}\bar{S}\right) ^{2}=0.
\end{equation}

On the other hand, the $M$ $\rightarrow \infty $ limit of the non-linear
gauge condition $A_{\mu }A^{\mu }=M^{2}$, is%
\begin{equation}
A_{0}\mp M=\pm \frac{A_{i}^{2}}{2M}.
\end{equation}%
The constant $M$ can be reabsorbed in $A_{0}$,\ so that $A_{0}\rightarrow
A_{0}\mp M$ and the gauge condition is
\begin{equation}
A_{0}=\pm \frac{A_{i}^{2}}{2M}.
\end{equation}%
In this way, the non relativistic gauge function $\lambda \;$is determined by%
\begin{equation}
\left( \tilde{\Phi}+\frac{1}{c}\partial _{t}\lambda \right) =-\frac{\left(
\tilde{A}^{i}-\partial _{i}\lambda \right) ^{2}}{2M},\;\;\;\;\tilde{A}^{\mu
}=(\tilde{\Phi},\mathbf{\tilde{A}}),  \label{NRLGCOND}
\end{equation}%
which is rewritten as%
\begin{equation}
\left( q\tilde{\Phi}+\frac{q}{c}\partial _{t}\lambda \right) =-\frac{\left(
\frac{q}{c}\tilde{A}^{i}-\frac{q}{c}\partial _{i}\lambda \right) ^{2}}{%
2\left( \frac{qM}{c^{2}}\right) }.  \label{NRGT}
\end{equation}%
Comparing (\ref{NRGT}) with (\ref{NONRELIMHJ}) we obtain%
\begin{equation}
-\frac{q}{c}\lambda =\bar{S},\;\;\;\;\;m=\frac{qM}{c^{2}},  \label{NRREL}
\end{equation}%
which reproduces the relativistic relation between\ $\Lambda \;$and$\;S.$

Let us consider the simple case of a constant magnetic field $\mathbf{B}=B%
\mathbf{\hat{k}}$. In the Coulomb gauge we have%
\begin{equation}
\tilde{\Phi}=0,\;\;\;\mathbf{\tilde{A}}=\frac{1}{2}\mathbf{B\times r.}
\end{equation}

Following Nambu's suggestion we use the Hamilton-Jacobi approach to
determine the gauge transformation function $\lambda $. To this end we
calculate the Hamilton-Jacobi principal function$\;\bar{S}$. As \ is well
known, this function corresponds to the action of the system evaluated at
fixed end points. We calculate%
\begin{equation}
\bar{S}(\mathbf{x}_{0},t_{0};\mathbf{x}_{1},t_{1})=\int_{t_{0}}^{t_{1}}dt\ L(%
\mathbf{x(}t\mathbf{),\dot{x}}(t)),
\end{equation}%
just by substituting the solution of the equations of motion into the
Lagrangian and integrating. We take the end point conditions as%
\begin{eqnarray}
x\left( t_{0}\right) &=&0,\;\;y\left( t_{0}=0\right) ,\;z\left( t_{0}\right)
=0,  \notag \\
x\left( t_{1}\right) &=&x_{1},\;y\left( t_{1}\right) =y_{1},\;z\left(
t_{1}\right) =z_{1}.
\end{eqnarray}%
The Lagrangian is%
\begin{equation}
L\left( \mathbf{x},\mathbf{\dot{x}},t\right) =\frac{m}{2}\mathbf{\dot{x}}%
^{2}+\frac{q}{c}\mathbf{\tilde{A}}\cdot \mathbf{\dot{x}}=\frac{m}{2}\mathbf{%
\dot{x}}^{2}+\frac{qB}{2c}\left( x\dot{y}-y\dot{x}\right) ,
\label{lagrangiano del B cte}
\end{equation}%
together with the equations of motion%
\begin{equation}
\ddot{x}-\frac{qB}{mc}\dot{y}=0\;,\;\;\;\ddot{y}+\frac{qB}{mc}\dot{x}%
=0\;,\;\;\;\;\ddot{z}=0.
\end{equation}%
The solutions are%
\begin{eqnarray}
x(t) &=&-\frac{1}{2}\left( x_{1}+y_{1}\cot \frac{\theta _{1}}{2}\right)
\left( \cos \omega t-1\right) +\frac{1}{2}\left( x_{1}\cot \frac{\theta _{1}%
}{2}-y_{1}\right) \sin \omega t, \\
y(t) &=&\frac{1}{2}\left( x_{1}+y_{1}\cot \frac{\theta _{1}}{2}\right) \sin
\omega t+\frac{1}{2}\left( x_{1}\cot \frac{\theta _{1}}{2}-y_{1}\right)
\left( \cos \omega t-1\right) , \\
z(t) &=&\frac{z_{1}}{t_{1}}t\;.
\end{eqnarray}%
with $\omega =qB/mc\;$and $\theta _{1}=\omega t_{1}$. Substituting them in
the Lagrangian we obtain%
\begin{equation}
L\left( \mathbf{x(t)},\mathbf{\dot{x}(t)}\right) =\frac{mz_{1}^{2}}{%
2t_{1}^{2}}+\frac{m\omega ^{2}}{8}\left( x_{1}^{2}+y_{1}^{2}\right) \csc
^{2}\left( \frac{\theta _{1}}{2}\right) \cos \omega t,
\end{equation}%
which produces the final result%
\begin{equation}
\bar{S}(\mathbf{x}_{0},t_{0};\mathbf{x}_{1},t_{1})=\frac{m}{2}\frac{z_{1}^{2}%
}{t_{1}}+\frac{m\omega }{4}\left( x_{1}^{2}+y_{1}^{2}\right) \cot \left(
\frac{\omega t_{1}}{2}\right) .
\end{equation}%
A direct calculation verifies that $\bar{S}\;$satisfies the Hamilton-Jacobi
equation:%
\begin{equation}
\frac{1}{2m}\left[ \left( \frac{\partial \bar{S}}{\partial x}+\frac{qBy}{2c}%
\right) ^{2}+\left( \frac{\partial \bar{S}}{\partial y}-\frac{qBx}{2c}%
\right) ^{2}+\left( \frac{\partial \bar{S}}{\partial z}\right) ^{2}\right] +%
\frac{\partial \bar{S}}{\partial t}=0.  \label{HJBCTE}
\end{equation}%
From Eq.(\ref{NRREL}) the gauge function results%
\begin{equation}
\lambda (\mathbf{x},t)=-\frac{1}{2}\frac{M}{c}\frac{z^{2}}{t}-\frac{B}{4}%
\left( x^{2}+y^{2}\right) \cot \left( \frac{Bc}{2M}t\right) .
\end{equation}%
Hence, the potentials in the non-linear gauge are%
\begin{eqnarray}
A_{0} &=&\frac{1}{c}\frac{\partial \lambda }{\partial t}= \frac{1}{2}\frac{M%
}{c^{2}}\frac{z^{2}}{t^{2}}+\frac{B^{2}}{8M}\left( x^{2}+y^{2}\right) \csc
\left( \frac{Bc}{2M}t\right) ,  \notag \\
A_{x} &=&\tilde{A}_{x}+\frac{\partial \lambda }{\partial x}=-\frac{B}{2}%
\left( y+x\cot \left( \frac{Bc}{2M}t\right) \right) ,  \notag \\
A_{y} &=&\tilde{A}_{y}+\frac{\partial \lambda }{\partial y}=+\frac{B}{2}%
\left( x-y\cot \left( \frac{Bc}{2M}t\right) \right) ,  \notag \\
A_{z} &=&\frac{\partial \lambda }{\partial z}=-\frac{M}{c}\frac{z}{t}.
\end{eqnarray}%
It is straightforward to verify that indeed we have%
\begin{equation}
A_{0}=\frac{1}{2M}\mathbf{A}^{2}.
\end{equation}

\section{The full SLSB model}

In this Appendix we explore the relation between electrodynamics and the
full SLSB model arising from the bumblebee model given in Eq. (\ref%
{LAGNAMBU0}). We emphasize again that this Lagrangian describes a physical
system only after the spontaneous symmetry breaking is realized. For
example, the mass parameter in Eq. (\ref{LAGNAMBU0}) has the wrong sign for
describing a massive vector field $B_{\mu }$ . As we have shown in Section
II, a correct realization is obtained only in the case $n^{2}<0$. We take $%
n^{2}=-1$ and the corresponding Lagrangian reads
\begin{eqnarray}
\mathcal{L(}A_{\mu },\omega ,\Theta \mathcal{)} &=&-\frac{1}{4}F_{\mu \nu
}F^{\mu \nu }\left( 1+\frac{\omega }{\sigma _{0}}\right) ^{2}+\frac{1}{%
\sigma _{0}}\left( 1+\frac{\omega }{\sigma _{0}}\right) F_{\mu \nu }\left(
A^{\mu }\partial ^{\nu }\omega \right)   \notag \\
&&+\frac{1}{2}\left( \partial _{\mu }\omega \right) \left( \partial ^{\mu
}\omega \right) +\frac{1}{2\sigma _{0}^{2}}A^{\nu }A^{\mu }\left( \partial
_{\mu }\omega \right) \left( \partial _{\nu }\omega \right)   \notag \\
&&-\frac{1}{2}\left( 2m^{2}\right) \omega ^{2}-\lambda \sigma _{0}\omega
^{3}-\frac{\lambda }{4}\omega ^{4}-J_{\mu }A^{\mu }\left( 1+\frac{\omega }{%
\sigma _{0}}\right) +\Theta \left( A_{\mu }A^{\mu }+\sigma _{0}^{2}\right) ,
\label{FULLSLSB}
\end{eqnarray}%
where we have introduced the non-linear condition via the Lagrange
multiplier $\Theta $. Here $\sigma _{0}^{2}=m^{2}/\lambda
,\;\;m^{2},\;\lambda >0.$

Next we examine the equations of motion in order to search for dynamical
conditions that could reproduce electrodynamics. In order to simplify the
dynamics it is convenient to adopt the following parametrization of the
Nambu-Goldstome modes $A_{\mu }$%
\begin{equation}
C_{\mu }=\left( 1+\frac{\omega }{\sigma _{0}}\right) A_{\mu }=\frac{\sigma }{%
\sigma _{0}}A_{\mu },\;\;\;C_{\mu \nu }=\partial _{\mu }C_{\nu }-\partial
_{\nu }C_{\mu },  \label{REDEFAMU}
\end{equation}%
where we also recall the notation $\sigma =\sigma _{0}+\omega $. This is
motivated precisely by the way in which the symmetry breaking was introduced
in Section II. As expected we have the identity%
\begin{equation}
-\frac{1}{4}C_{\mu \nu }C^{\mu \nu }=-\frac{1}{4}\frac{\sigma ^{2}}{\sigma
_{0}^{2}}F_{\mu \nu }F^{\mu \nu }+\frac{\sigma }{\sigma _{0}^{2}}F_{\nu \mu
}A^{\nu }\partial ^{\mu }\omega +\frac{1}{2}\partial _{\mu }\omega \partial
^{\mu }\omega +\frac{1}{2\sigma _{0}^{2}}A^{\nu }A^{\mu }\partial _{\mu
}\omega \partial _{\nu }\omega ,  \label{IDENTITY}
\end{equation}%
which reproduces de first two lines on the right hand side of (\ref{FULLSLSB}%
). The result is\textbf{\ }%
\begin{equation}
\mathcal{L}(C_{\mu },\omega ,\Omega )=-\frac{1}{4}C_{\mu \nu }C^{\mu \nu
}-m^{2}\omega ^{2}-\lambda \sigma _{0}\omega ^{3}-\frac{\lambda }{4}\omega
^{4}-J_{\mu }C^{\mu }+\frac{1}{2}\Omega \left( C_{\mu }C^{\mu }+\left(
\sigma _{0}+\omega \right) ^{2}\right) ,\;\;\;  \label{FINLAGREP}
\end{equation}%
with a redefinition of the Lagrange multiplier $\Theta $. The equations of
motion are\textbf{\ }%
\begin{eqnarray}
\partial _{\mu }C^{\mu \nu }-J^{\nu }+\Omega C^{\nu } &=&0,  \label{EQBB1} \\
C_{\mu }C^{\mu }+\left( \sigma _{0}+\omega \right) ^{2} &=&0,  \label{EQBB2}
\\
-2m^{2}\omega -3\lambda \sigma _{0}\omega ^{2}-\lambda \omega ^{3}+\Omega
\left( \sigma _{0}+\omega \right)  &=&0,  \label{EQBB3}
\end{eqnarray}%
where the last one  can be rewritten as
\begin{equation}
\left( \sigma _{0}+\omega \right) \left[ \Omega -\lambda \omega \left(
\omega +2\sigma _{0}\right) \right] =0.  \label{EQBB4}
\end{equation}%
This equation is satisfied for arbitrary $\Omega \;$when $\omega =-\sigma
_{0}$. In this case we have
\begin{equation}
C_{\mu }=\left( 1+\frac{\omega }{\sigma _{0}}\right) A_{\mu }=0
\end{equation}%
and the remaining \ equations consistently lead to $J^{\nu }=0$, so that the
Goldstone field has no dynamics. The physically relevant configurations
arise when $\omega \neq -\sigma _{0}$. Then Eq. (\ref{EQBB4}) reads%
\begin{equation}
\Omega -\lambda \omega \left( \omega +2\sigma _{0}\right) =0.  \label{EQBBM5}
\end{equation}%
Following the discussion in subsection C-1 in Ref. \cite{BLUHM}\ we could
explore the possibility of setting the initial condition $\Omega (x,t=0)=0$.
Notice that this would require to set the four Maxwell equations, including
the Gauss law, as initial conditions, according to Eq.(\ref{EQBB1}). We
prefer to think the other way round and we impose $\left( \partial
_{i}C^{i0}-J^{0}\right) =0$ as an initial condition. Since Eq.(\ref{EQBB1})
has the same form as Eq.(\ref{a00}), the dynamics guarantees that the Gauss
law with respect to the field $C^{i0}$ is valid for all times, which in turn
leads to $\Omega (x,t)=0$, from the zero component of Eq.(\ref{EQBB1}). That
is to say, we now have%
\begin{equation}
\left( \partial _{i}C^{i0}-J^{0}\right) (\mathbf{x},t)=0\ \ \ \ \rightarrow
\;\Omega (\mathbf{x},t)=0.
\end{equation}%
In this way we also recover Maxwell equations for $C^{\mu \nu
}(x,t):\;\partial _{\mu }C^{\mu \nu }-J^{\nu }=0$. \ Nevertheless, let us
observe that we are interested in the Gauss law $\left( \partial
_{i}F^{i0}-J^{0}\right) =0$ \ together with Maxwell equations for $F_{\mu
\nu }$.$\;$To this end we consider the equation of motion (\ref{EQBBM5})
which yields the solutions\textbf{\ }
\begin{align}
\omega & =0,\;\;\;\;\rightarrow \;\;\;C_{\mu }=A_{\mu }\ ,\ \ \;\;\;\partial
_{\mu }F^{\mu \nu }=J^{\nu }\ .\ \ \ \ \ \  \\
\ \ \ \omega & =-2\sigma _{0},\;\;\rightarrow \;C_{\mu }=-A_{\mu
},\;\;\;\partial _{\mu }F^{\mu \nu }=-J^{\nu }.
\end{align}%
In both situations the gauge fixing condition%
\begin{equation}
A_{\mu }A^{\mu }+\sigma _{0}^{2}=0  \label{NLGC}
\end{equation}%
is satisfied.\ The case $\omega =0$\ describes the electromagnetic field, in
the non-linear gauge (\ref{NLGC}), coupled to the current $J^{\nu }$, while $%
\omega =-2\sigma _{0}$\ corresponds to an electromagnetic field in the same
gauge, but coupled to the current $-J^{\nu }$.

$\;$Summarizing, we have shown that the imposition of the Gauss law as an
initial condition in the full spontaneously broken space-like bumblebee
model leads to electrodynamics in the non-linear gauge$\;A_{\mu }A^{\mu
}+\sigma _{0}^{2}=0.$\ The situation is analogous to the Nambu model, except
that here we have obtained the condition $\omega (x,t)=0$ dynamically as the
consequence of imposing the Gauss as an initial condition, together with
satisfying the corresponding equations of motion.

\section*{Acknowledgements}

L.F.U is partially supported by projects CONACYT \# 55310 and
DGAPA-UNAM-IN111210. He also acknowledges support from RED-FAE, CONACYT.
R.M. acknowledges support from CONICET-Argentina. O.J.F. was partially
supported by the project CONACYT \# 55310.


\begin{thebibliography}{99}
\bibitem{PRLNielsen} J.L, Chkareuli, C.D. Froggat and H.B. Nielsen, Phys.
Rev. Letts. \textbf{87, }091601\ (2001).

\bibitem{Nambu} Y. Nambu, Suppl. of the Prog. Theor. Phys., Extra Number,
190\ (1968).

\bibitem{Bjorken} J. D. Bjorken, Ann. Phys. (N.Y.) \textbf{24, }174 (1963).

\bibitem{others} G. S. Guralnik, Phys. Rev. \textbf{134, }B 1404\ (1964); R.
Righi and G. Venturi, Il Nuovo Cimento \textbf{A43}, 145\ (1978); G.
Venturi, Il Nuovo Cimento \textbf{A63},\ 64\ (1981); R. Righi and G.
Venturi, Int. J. of Theor. Phys. \textbf{21},\ 63\ (1982); A. Kovner and B.
Rosenstein, Phys. Rev. \textbf{D49}, 5571\ (1994); I. Low and A. V. Manohar,
Phys. Rev. Lett. \textbf{88},101602\ (2002); P. Kraus and E. T. Tomboulis,
Phys. Rev. \textbf{D66}, 045015\ (2002).

\bibitem{kostelecky1} V.A. Kostelecky and R. Potting, Gen.\ Rel.\ Grav.\
\textbf{37}, 1675\ (2005), Int.\ J.\ Mod.\ Phys. \textbf{D14}, 2341\ (2005).

\bibitem{chkareulli} J.L. Chkareuli, C.D. Froggat and H.B. Nielsen, Nucl.
Phys. \textbf{B821}, 65 (2009).

\bibitem{azatov} A. T. Azatov and J.L. Chkareuli, Phys. Rev. \textbf{D73},
065026 (2006).

\bibitem{mohapatra} J.L. Chkareuli, C.D. Froggat, R. N. Mohapatra and H.B.
Nielsen, \textit{Photon as a Vector Goldstone Boson: Nonlinear $\sigma$
Model for QED}, arXiv: hep-th/0412225.

\bibitem{chkareulli1} J.L. Chkareuli and J. G. Jejelava, Phys. Lett. \textbf{%
B659}, 754 (2008).

\bibitem{BLUHM} R. Bluhm, N. L. Gagne, R. Potting and A. Vrublevskis, Phys.
Rev. \textbf{D77}, 125007 (2008).

\bibitem{kostelecky2} R. Bluhm and V. A. Kostelecky, Phys. Rev.\textbf{D71},
065008\ (2005).

\bibitem{Diracfull} P.A.M. Dirac, Proc. Roy. Soc. (London), \textbf{A209},\
291\ (1951); P.A.M. Dirac, Proc. Roy. Soc. (London), \textbf{A212},\ 330\
(19512); P.A.M. Dirac, Proc. Roy. Soc. (London), \textbf{A223},\textbf{\ }%
438\ (1954).

\bibitem{Dirac1} P.A.M. Dirac,\textit{\ }Nature (London)\textbf{168},\ 906\
(1951).

\bibitem{joglekar} S.D. Joglekar, Pramana-J. Phys. \textbf{32},\ 195\ (1989).

\bibitem{NLSIGMAM} See for example S. Weinberg,\textit{The Quantum Theory of
Fields, Vol. II}, Cambridge University Press, New York, 1996.
\end{thebibliography}
\end{document}